\documentclass[twocolumn,preprintnumbers,amsmath,amssymb]{revtex4-2}

\usepackage{graphicx}  
\usepackage{sidecap}
\begin{document}
%
\title{Transit-time resonances enabling  amplification and generation of terahertz radiation in periodic graphene p-i-n structures with the Zener-Klein interband tunneling
}
\author{V.~Ryzhii$^{1}$, M.~Ryzhii$^{2}$, 
V.~Mitin$^3$, M.~S.~Shur$^4$, and  T.~Otsuji$^1$}
\address{
$^1$Research Institute of Electrical Communication,~Tohoku University,~Sendai~980-8577, Japan\\
$^2$Department of Computer Science and Engineering, University of Aizu, Aizu-Wakamatsu 965-8580, Japan\\
$^3$Department of Electrical Engineering, University at Buffalo, SUNY, Buffalo, New York 14260 USA\\
$^4$Department of Electrical,~Computer,~and~Systems~Engineering, Rensselaer Polytechnic Institute,~Troy,~New York~12180,~USA
}

\begin{abstract} 
\normalsize
The Zener-Klein (ZK) interband tunneling in graphene layers (GLs) with the lateral n-i-n and p-i-n junctions results in the nonlinear I-V characteristics that can be used for the rectification and detection of the terahertz (THz) signals. The transit time delay of the tunneling electrons and holes in the depletion regions leads to the phase shift between the THz current and THz voltage causing the negative dynamic conductance in a certain frequency range and 
resulting in the so-called transit-time (TT) instability. The combination of the ZK tunneling and the TT negative dynamic conductance enables resonant THz detection and the amplification and generation of THz radiation. We propose and evaluate the THz devices based on periodic cascade GL p-i-n structures exhibiting the TT resonances (GPIN-TTDs). Such structures can serve as THz amplifiers and, being placed in a Fabri-Perot cavity, or coupled to a THz antenna or using a ring oscillator connection, as THz radiation sources. 
\end{abstract} 
\maketitle
\section{Introduction}

Since the prediction of the Dyakonov-Shur plasma instability in the current-driven two-dimensional  electron channels of the field-effect transistor  structures leading to
the terahertz (THz) emission~\cite{1,2},  such an emission have been 
extensively discussed in the literature (see, for example,~\cite{3,4,5,6,7,8,9,10,11,12,13,14,15} and the references therein).
As demonstrated recently~\cite{16,17,18}, the Coulomb electron drag~\cite{19,20,21,22,23} by the injected hot electrons or/and holes
in the graphene layer (GL) n-i-n and p-i-n structures~\cite{1,2,15,24,25} (see also~\cite{26}), 
could enable a specific instability mechanism. In these structures, the gated regions play the role of the plasmonic resonant cavities.
The finite transit time  of the holes and electrons injected or generated in the depleted regions of the GL channel can enable the negative dynamic conductivity  on the order of   the inverse
transit time~\cite{27,28}.

In this paper, we propose and analyze   the periodic ungated  GL p-i-n transit time (TT) device (GPIN-TTD) with the i-region exhibiting the negative dynamic conductivity due to the TT effect.
The generation of the holes and electrons in the i-region is associated with the Zener-Klein (ZK) interband tunneling~\cite{29,30,31,32}. 
In contrast to our previous works on the TT effects  in GL structures, here we assume the following:\\

-- The GPIN-TTDs are based on periodic structures comprising  the alternating reverse-biased p-i-n junctions with the depleted i-region and  the forward-based junctions. 
The resistance of the forward-based junctions junctions can be of the order of or even exceed the resistance of the doped p- and n-regions, despite the forward bias;\\

-- The ungated d-regions are  chemically doped. The plasma (PL) frequency of these regions
is markedly larger than that of the gated regions and  
the ungated PL resonances are out of the frequency range under consideration;\\

-- The Coulomb carrier drag in the relatively long d-regions  in the GPIN-TTDs studied in~\cite{19,20,21,22,23}
is weak due to a short momentum relaxation time of the carriers injected into these regions from the i-region,
especially at  elevated voltage biases when the emission of optical phonons is essential.\\

-- The combination of  the ZK interband tunneling and the TT effects, which, in particular,  enables  
 the rectification and detection of the impinging THz radiation~\cite{33}, might provide the THz radiation amplification and generation.

As we show below, the TT effects in the GPIN-TTD can result in the negative real part of its impedance enabling
the amplification of the THz radiation. In the GPIN-TTDs with the optimized parameters, the TT resonances might be associated with the interplay of the i-region capacitance and the d-regions inductance.
This can lead to the self-excitation of the THz oscillation and generation of the THz radiation in the GPIN-TTDs supplied by a proper antenna or incorporated into a ring oscillator configuration. 
Since the overall dimension of the periodic structure could
be comparable to the THz radiation wavelength, it could also
take a role of a distributive coupling antenna structure. Alternatively, the GPIN-TTDs can serve as an active media in the THz sources using   Fabri-Perot cavities.

\begin{figure*}\centering
\includegraphics[width=13.0cm]{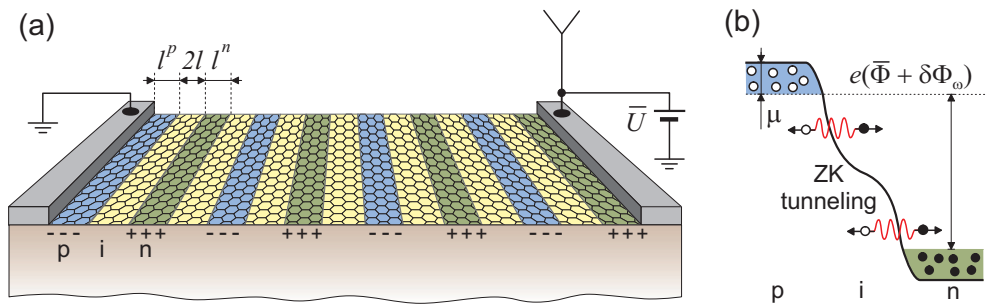}
\caption{(a)
Schematic view of   periodic GPIN-TTD structure  with  chemically doped  p- and n-regions of the GL with a THz antenna  and (b) potential profile of the reverse-biased p-i-n junction in one of the sections   (periods). Symbols "+" and "-"
correspond to donors and acceptors, and 
open and opaque circles represent holes and electrons, respectively.} 
\label{F1}
\end{figure*}

\section{Device model}

We consider the periodic GPIN-TTD structure based on a GL  channel embedded in a dielectric, for example, hBN or polyimide, or placed on a dielectric  substrate and having  a thin top passivating layer.
The GPIN-TTD structure can also be  suspended in air or vacuum.
 The GL channel comprises a periodic array of the undoped i-regions of the length $2l$  placed between the chemically doped  p- and n-regions  (d-regions)  of the lengths $l^p$ and $l^n$ ($l^p=l^n = L$).
The channel is bounded by 
the side source and drain contacts. The drain-to-source bias voltage ${\overline U}$
 results in 
alternating  lateral reverse biased p-i-n  junctions (with the depleted i-regions), p- and n-quasi-neutral
region,  and the forward biased n-i-p junctions 
(with the i-regions in these junctions, which below are referred to as the f-regions, filled with the injected electrons and holes and having a high conductance).

The bias voltage, ${\overline V}$,
per one period (section) of the structure (assuming that the sections are equivalent) is equal to 
${\overline V} = {\overline U}/N$, 
where $N$ is the number of  sections.
Figure~1  shows the GPIN-TTD structure (the top passivating layer is not shown) and the potential profile of the p-i-n junction in one of the reverse-biased device sections. 
The GPIN-TTD  can be connected with a THz antenna, can serve as a distributed antenna,  form a ring oscillator structure,
or can be  placed in the Fabri-Perot cavity.

We assume that the holes  and electrons  generated in the i-regions due to the ZK interband tunneling  propagate ballistically or near ballistically~\cite{34,35,36,37,38,39,40}. 
Due to the specific of the ZK interband tunneling in GLs~\cite{29}, the velocity, $v = \pm v_W$, of the generated electrons and holes is directed primarily along the electric field in the
i-region, i.e., in the $x-$ direction. 
If the i-region length  $2l \lesssim 1~\mu$m, the transport of the injected holes and electrons  in the  i-region 
can be ballistic 
 at  room and lower temperatures~\cite{40,41}.
 In this case, the average electron and hole velocities across the entire i-region
are   equal to $\pm v_W$, where $v_W \simeq 10^8$~cm/s is the characteristic carrier velocity in GLs.
%


%
If the voltage drop across the i-regions between  the  d-regions $\Phi$ includes the AC component $\delta \Phi_{\omega}$, i.e.,  $\Phi = {\overline \Phi}
+ \delta \Phi_{\omega}\exp(-i\omega\,t)$ (where $\delta \Phi_{\omega}$ and $\omega$ are the THz signal amplitude and frequency),
the  holes and electrons generated due to tunneling and  propagated in the i-regions  
  (and the GL channel as a whole) induce
the terminal AC current. Due to the finite time of the electron transit, the induced AC current and the AC voltage could have the opposite phase. The latter implies that the dynamic conductance of the i-region can be negative. 

The impedance of the periodic GPIN-TTD structure $Z_{\omega}^{GPIN-TTD}$ is the sum of the impedances of the sections $Z_{\omega}$, i.e., $Z_{\omega}^{GPIN-TTD} = NZ_{\omega}$  with

\begin{eqnarray}\label{eq1}
Z_{\omega} = Z_{\omega}^p +Z_{\omega}^i+Z_{\omega}^n + Z_{\omega}^f,
\end{eqnarray}
 where  $Z_{\omega}^p$, $Z_{\omega}^i$,  $Z_{\omega}^n$, and $Z^f_{\omega}$  are the impedances of the pertinent regions.
 For $N$ equivalent sections $Z_{\omega}^{GPIN-TTD} = NZ_{\omega}$.  

Below we calculate the frequency-dependent impedance of the  GPIN-TTD  as a function of the structural characteristics. 
The numerical calculations were performed using Maple 2021 software (Maplesoft, Waterloo, ON, Canada).

\begin{figure}[b] \centering
\includegraphics[width=6.5cm]{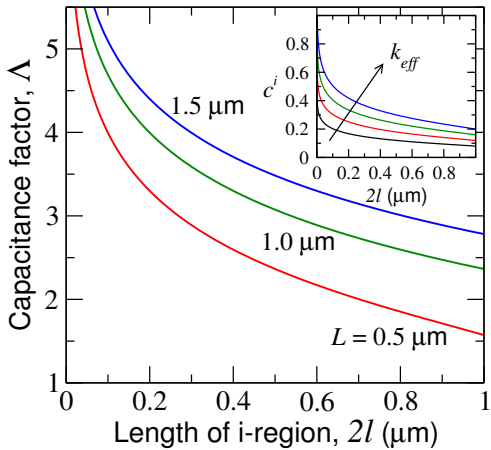}
\caption{
  Capacitance factor $\Lambda$ versus i-regions length $2l$ for different d-regions length $L$
  and i-region capacitance $c^i$ (inset) for $L = 1.0~\mu$m and different values of dielectric constant $\kappa_{eff} = 1,\, 1.5,\, 2,$ and 2.5.
} 
\label{F2}
\end{figure}

\begin{figure}[t] \centering
\includegraphics[width=6.5cm]{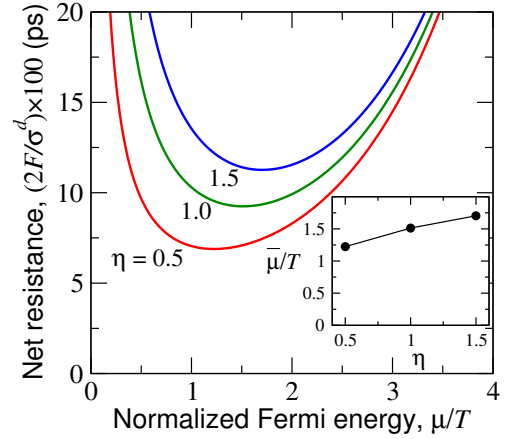}
\caption{ Net DC resistance of highly conducting regions $2F/\sigma_d$ versus normalized Fermi energy $\mu/T$
for different values of parameter $\eta = \nu\,L/v_W$ ($T = 300$~K).  Inset shows  ${\overline \mu}/T$ as a function of parameter $\eta$.
} 
\label{F3}
\end{figure}
 
\begin{figure*}\centering
\includegraphics[width=13.0cm]{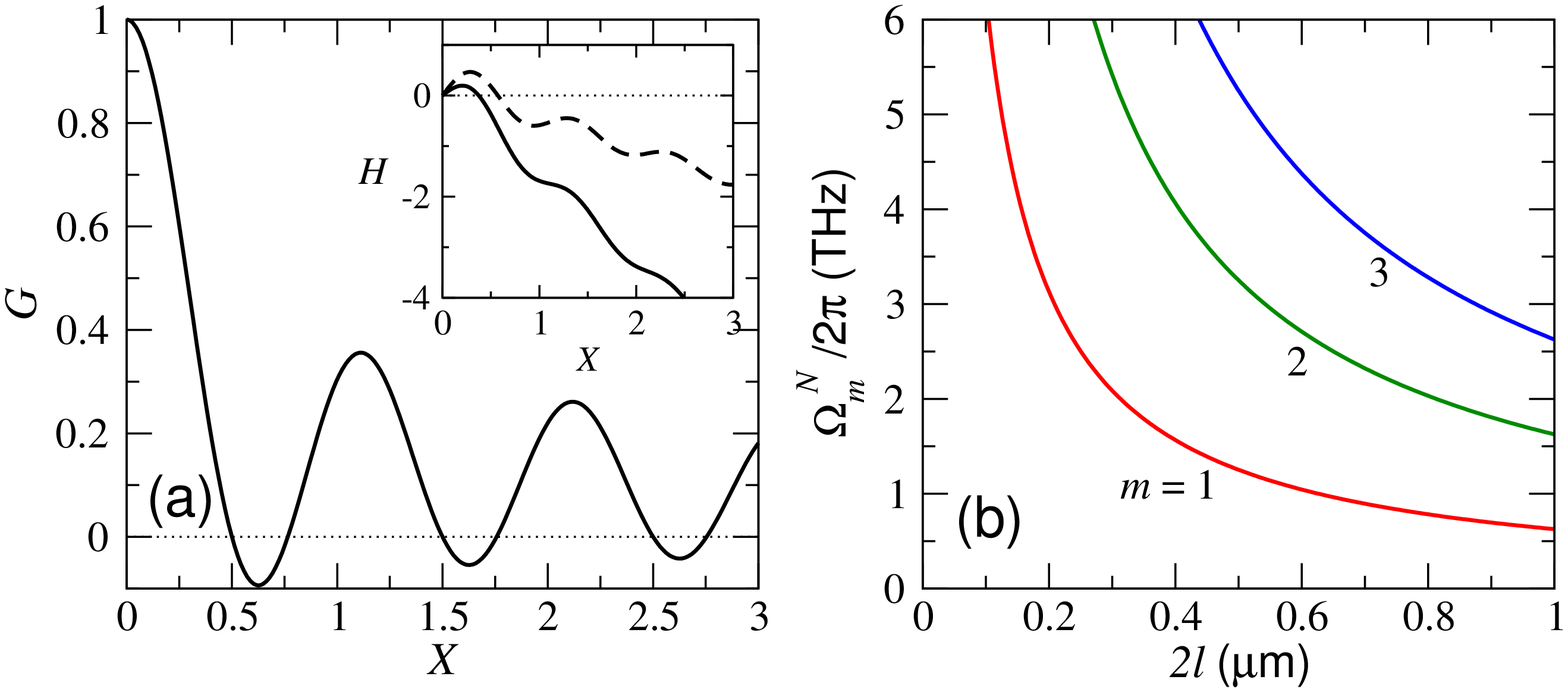}
\caption{(a) Function $G(\pi\,X)$ and  
 $H(\pi\,X)$ (inset)  for $L = 1.0~\mu$m, ${\overline \Phi} = 100$~mV, and $2l = 0.25~\mu$m -solid line and $2l = 0.75~\mu$m -dashed line; (b) Characteristic TT frequencies $\Omega_m^N/2\pi = X_m v_W/2l$, 
 corresponding to the function $G(\pi\,X)$ minima, versus the i-region length $2l$.
} 
\label{F4ab}
\end{figure*}

\section{The impedance of the GPIN-TTD}

 For the most interesting voltage range $e{\overline \Phi}_i > T$, 
 where $T$ is the temperature in the energy units, 
 we obtain for the AC current density (the current per unit width of the device in the direction along the gate edges)

\begin{eqnarray}\label{eq2}
\delta J_{\omega } = \sigma_{\omega}^i\delta \Phi_{\omega}.
\end{eqnarray}
For  for  the blade-like p- and n-regions  and the electric field concentrated near these regions~\cite{27},
the AC conductance accounting for  the 
frequency-dependent factor associated with the TT effect (with the finiteness of the hole and electron TT, $t^i = 2l/v_W$, across the i-region) is given by

\begin{eqnarray}\label{eq3}
\sigma_{\omega}^i = \sigma^i \displaystyle e^{i\omega\,t^i/2}{\mathcal J}_0(\omega\,t^i/2)  -i\omega\,c^i.  
\end{eqnarray}
The differential DC  conductance of the i-region~\cite{27,28,33}, determined by the DC voltage drop, ${\overline \Phi}$, across this region,
and  the geometrical capacitance of this region are
\begin{eqnarray}\label{eq4}
\sigma^i= b\frac{e^2}{\hbar}\sqrt{\frac{e{\overline \Phi}}{2l\hbar\,v_W}}
\end{eqnarray} 
 and

\begin{eqnarray}\label{eq5}
c^i= \frac{\kappa_{eff}}{2\pi^2}\Lambda \propto \kappa_{eff},
\end{eqnarray} 
respectively.
In Eqs.~(4) and (5), $e = |e|$ is the carrier charge, $b= [3\Gamma(1/4)\Gamma(1/2)/2\Gamma(3/4)]\simeq 0.0716$~\cite{28},  $\kappa_{eff}$
is the effective dielectric constant accounting for  the dielectric constant, $\kappa_S$ and $\kappa_T$, of  the substrate and top/passivation layer at the frequency in the THz range and their thickness, $\Lambda = (L/l)\tan^{-1}\frac{1}{\displaystyle\sqrt{(L/l)^2-1}}+
\ln[(L/l)+ \sqrt{(L/l)^2-1}] \simeq \ln(2L/l)$~\cite{41} (see also~\cite{42,43,44,45}) is a logarithmic factor depending on the device geometrical parameters (capacitance factor), 
and
 ${\mathcal J}_0(s)$ is the Bessel functions. Generally, $\kappa_{eff} = (\kappa_S + \kappa_T)/2$. In the case of a rather thin top/passivation layer with the thickness $w \ll 2l$, $\kappa_{eff} \simeq (\kappa_S + 1)/2)$. 
The factor of 1/2 is because the streamlines of electric field emerge out of the top layer.

Examples of  $\Lambda$ and $c^i$ as functions of  the i-region length $2l$ are shown in Fig.~2.

The Drude formula yields the following expression for the AC conductance of the d-region:

\begin{eqnarray}\label{eq6}
\sigma_{\omega}^d  
= \frac{i\nu}{\omega +i\nu}\sigma^d,
\end{eqnarray}
where 
\begin{eqnarray}\label{eq7}
\sigma^d =  \frac{e^2\mu}{\pi\hbar^2L\nu} \propto \frac{\mu}{L\nu}
\end{eqnarray}
is the doped region DC conductance, $\nu$ is the collision frequency of the carriers in the d-regions with
phonons, impurities (remote and residual), and defects, and $\mu$ is the carrier Fermi energy in these regions. Equation~(7) accounts for the kinetic inductance associated with the quasi-equilibrium holes and electrons in the d-region.

Considering Eqs.~(6) and  (7) and accounting for the d-regions capacitance, we obtain 
 the d-region  impedance: 

\begin{eqnarray}\label{eq8}
Z_{\omega}^{d}
\simeq \frac{1}{\displaystyle \sigma^d\frac{\nu^2}{(\omega^2 + \nu^2)} +i\biggl[\sigma^d\nu\frac{\omega}{(\omega^2 + \nu^2)} -\omega\,c^d\biggr]}\nonumber\\
 =  \frac{1}{\displaystyle \sigma^d\frac{\nu^2}{(\omega^2 + \nu^2)} +i\sigma^d\nu\biggl[\frac{\omega}{(\omega^2 + \nu^2)} -\omega\,c^d{\mathcal L}^d\biggr]}\nonumber\\
=\frac{1}{\sigma^d}\frac{(\omega^2 + \nu^2)}{\nu^2}\frac{1}{1 +i\displaystyle\frac{\omega}{\nu}\biggl[\frac{\Omega_{PL}^2 -(\omega^2+\nu^2)}{\Omega_{PL}^2}\biggr]} .
\end{eqnarray}
Here  ${\mathcal L}^d =  1/\sigma^d\nu$ and $c^d$ are the d-region inductance and capacitance, respectively, and

\begin{eqnarray}\label{eq9}
\Omega_{PL} = \frac{1}{\sqrt{c^d{\mathcal L}^d}} =\sqrt{\frac{4\pi\,e^2\mu}{\kappa_{eff}\hbar^2L}} 
\end{eqnarray}
is the PL frequency for the {\it ungated} d-regions of the GL channel.  
Since $c^d$ is  small, the   frequency $\Omega_{PL}$ can be fairly large 
compared to the gated graphene structures with the same doping and length due to a relatively small capacitance of the ungated d-regions.

In the following, we assume that  the signal frequency $\nu < \omega \ll \Omega_{PL}$.
In this case, accounting for the f-region impedance, for
the net impedance of the highly conducting regions (the d- and f-regions)  in the periodic structure section we obtain:
(see Appendix A)

\begin{eqnarray}\label{eq10}
2Z_{\omega}^{d} + Z_{\omega}^f
\simeq \frac{2}{\sigma^d}F
 - i\frac{2}{\sigma^d}\frac{\omega}{\nu}.
\end{eqnarray}
Here the factor 

\begin{eqnarray}\label{eq11}
F =
1 + \biggl(\frac{\pi\,v_W}{8L\nu}\biggr)\frac{\mu}{(\mu\ln2 + \pi^2T/12)}\exp\biggl(\frac{\mu}{T}\biggr)
\end{eqnarray}
reflects the contribution of the f-regions to the  GL channel
resistance. In Eq.~(10), we disregarded the reactive component of the f-region conductance.

Accounting for Eqs.~(1)  and (10), we obtain 
 the net impedance, $Z_{\omega}$,  of one 
 of the periodic structure section 
 
\begin{eqnarray}\label{eq12}
Z_{\omega} = \frac{2}{\sigma^d}F -
i\frac{2}{\sigma^d}\frac{\omega}{\nu} +
\frac{1}{\sigma^i e^{i\omega\,t^i/2}{\mathcal J}_0(\omega\,t_i/2) -i\omega\,c^i}.
\end{eqnarray}

In the low-frequency limit, we obtain from Eq.~(12) $Z_{0} = 1/\sigma^i +2F/\sigma^d$.

The quantity $2F/\sigma^d$ exhibits a minimum as  a function of $\mu$  at
$\mu = {\overline \mu} \sim T$ with ${\overline \mu}$ determined by the  parameter $\eta =(\nu\,L/v_W)$.
For $L = (0.5 - 1.5)~\mu$m and $\nu = 1$~ps, one obtains $\eta \simeq 0.5 - 1.5$. 
Figure 3 shows the dependence of the net DC resistance of the highly conducting parts of the channel (comprising the d- and f-regions) on the  normalized Fermi energy $\mu/T$ calculated for different values of the parameter $\eta$. The  inset shows 
 the Fermi energy, ${\overline \mu}$, corresponding to the net resistance minimum,  as a function of $\nu$.

In the following, 
we chose $\mu = {\overline \mu}$. Table I provides  min~($2F/\sigma^d$) and min~($2/\sigma^d$)
calculated for several values of the parameter $\eta$. One can see from Table I that the resistance of the f-regions is crucial because  min~$(2F/\sigma^d) > {\rm min}(2/\sigma^d)$.

\begin{table}[t]
\caption{\label{table} GPIN-TTD parameters ($T = 300$~K)}  
\vspace{2 mm}
\centerline{\begin{tabular}{lcccc}
\hline\hline
&  $\eta =\nu\,L/v_W$ &${\overline \mu}/T$\,  &min($2F/\sigma^d)\times 100$~ps\,  &  min($2/\sigma^d)\times 100$~ps 
\\ \hline \hline
 &0.25 &0.98  &5.11  &1.235 \\
 &0.5 &1.22  &6.89  &2.66 \\
 &0.1 &0.72  &3.98  &0.91  \\
 &1.0  &1.51  &9.23  &4.30 \\
 &1.5  &1.70  &11.26 &5.74 \\                                                                                                                                                                                                                                                                                                                                                                                                                                                                                                                                                                                                                                                                                                                                                                                                                                                                                                                                                                                                                                       
\hline\hline
\end{tabular}}
\end{table}

\begin{figure*}[t] \centering
\includegraphics[width=13.0cm]{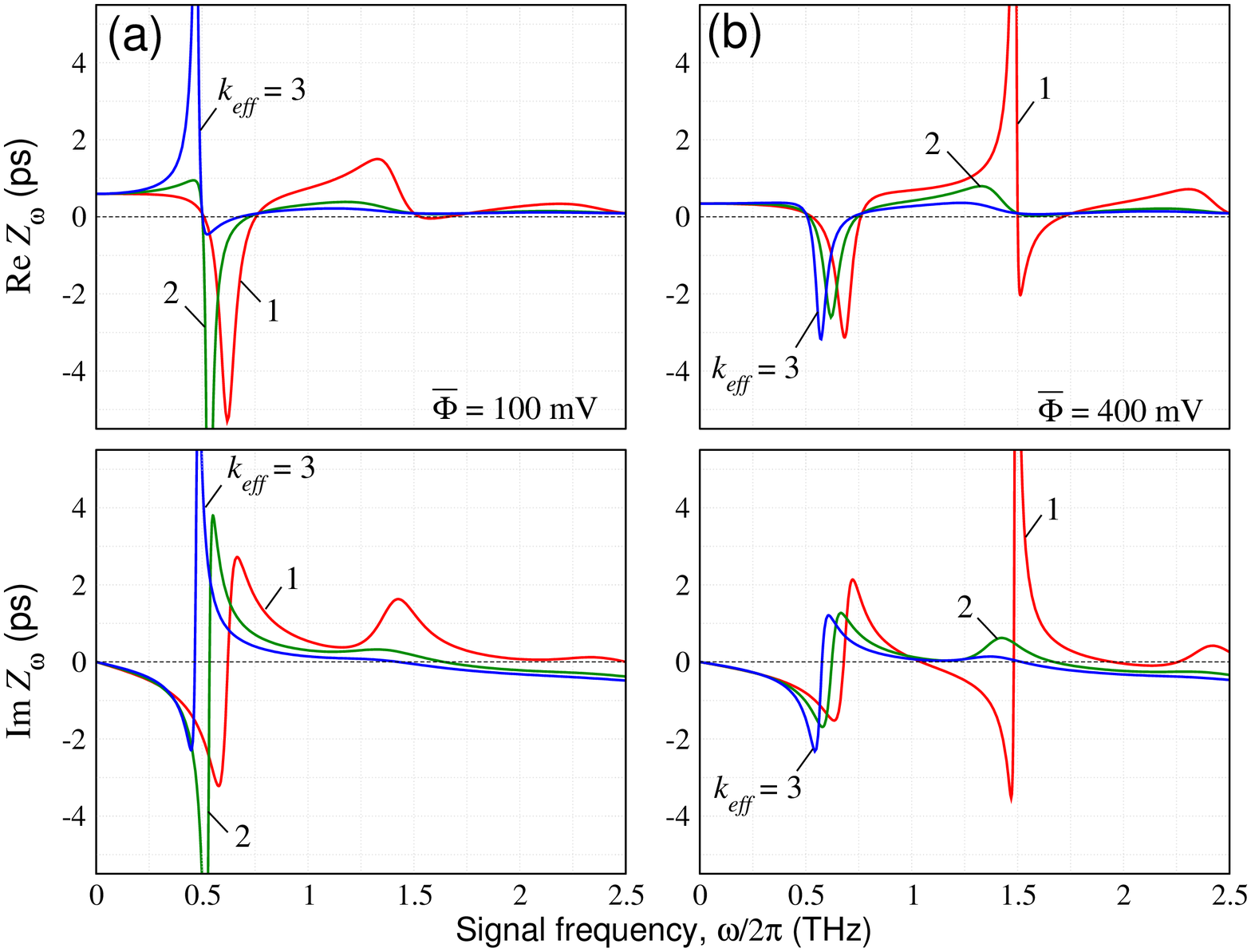}
\caption{The frequency dependences of the real and imaginary parts of the impedance, 
Re~$Z_{\omega}$ and Im~$Z_{\omega}$, of
the GPIN-TTD with $2l = 1.0~\mu$m and $L = 1.0~\mu$m
 for different $k_{eff}$ at (a) ${\overline \Phi} = 100$~mV and (b)
  ${\overline \Phi} = 400$~mV ($\Omega_1^N/2\pi \simeq 0.63$~THz).
} 
\label{F5ab}
\end{figure*}

\begin{figure*}[t] \centering
\includegraphics[width=13.0cm]{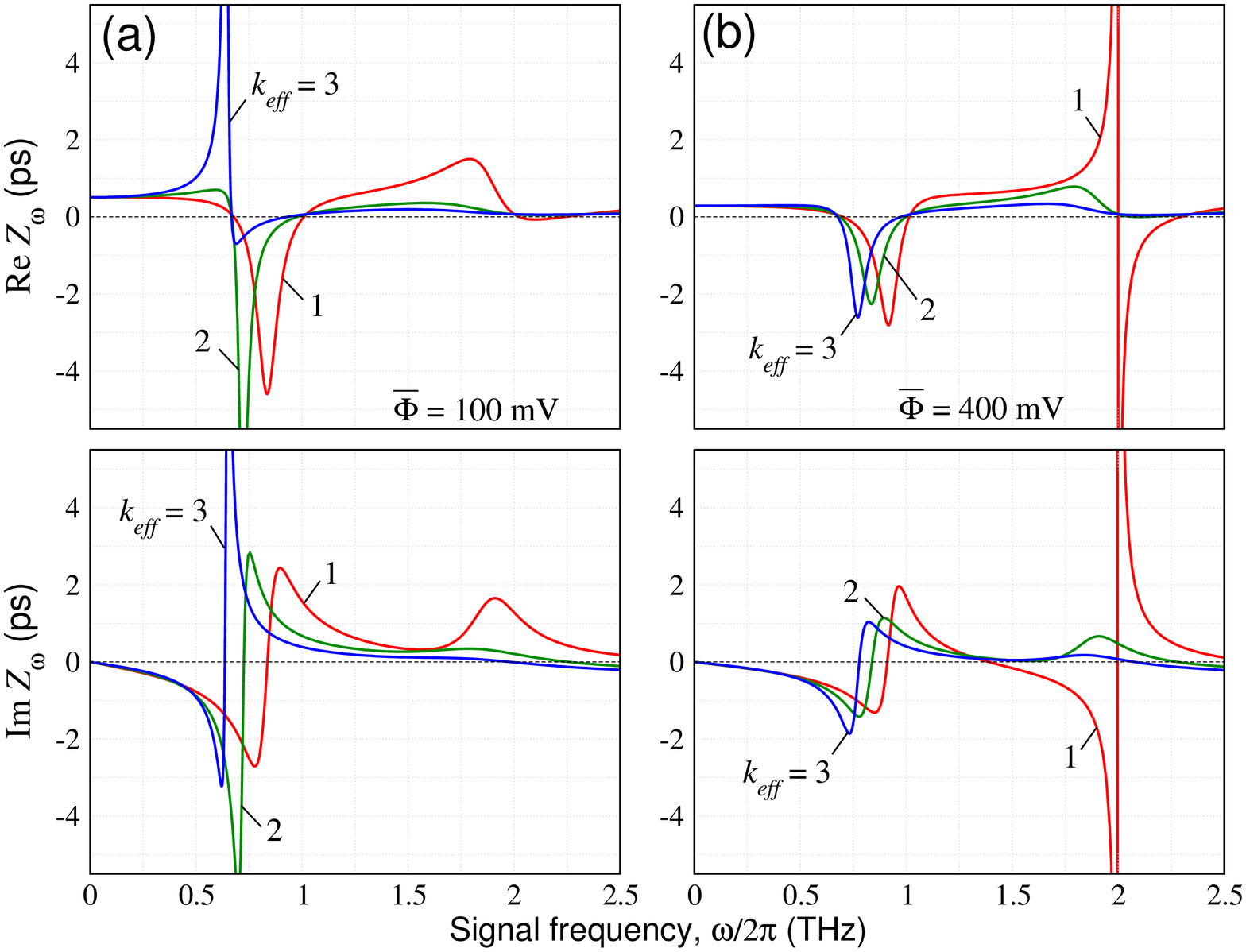}
\caption{The frequency dependences of the real and imaginary parts of the impedance, 
Re~$Z_{\omega}$ and Im~$Z_{\omega}$, of
the GPIN-TTD with $2l = 0.75~\mu$m and $L = 0.5~\mu$m
 for different $k_{eff}$ at (a) ${\overline \Phi} = 100$~mV and (b)
  ${\overline \Phi} = 400$~mV ($\Omega_1^N/2\pi  \simeq 0.84$~THz).
}  
\label{F6ab}
\end{figure*}

For the real and imaginary parts of the GPIN-TTD section  Eq.~(12) yields:

\begin{eqnarray}\label{eq13}
{\rm Re}Z_{\omega} \simeq  \frac{2}{\sigma^d}F
+\frac{G(\omega\,t^i/2)}
{\sigma^i[G^2(\omega\,t^i/2) + H^2(\omega\,t^i/2)]},
\end{eqnarray}

 \begin{eqnarray}\label{eq14}
{\rm Im} Z_{\omega} \simeq- \frac{2}{\sigma^d}\frac{\omega}{\nu}
- \frac{H(\omega\,t^i/2)}
{\sigma^i[G^2(\omega\,t^i/2) +H^2(\omega\,t^i/2)]} .
\end{eqnarray}
Here
 \begin{eqnarray}\label{eq15}
G(\pi\,X) = {\mathcal J}_0(\pi\,X)\cos(\pi\,X),
\end{eqnarray}

\begin{eqnarray}\label{eq16}
H(\pi\,X) = {\mathcal J}_0(\pi\,X)\sin(\pi\,X)
-\pi\,X\xi^i,
\end{eqnarray}
where  $X = \omega\,t^i/2\pi$ and $\xi^i =(2c^i/t^i\sigma^i)$.
 The first terms on the right-side of Eqs.~(13) and (14) reflect
the resistance and kinetic induction of the highly conducting regions, respectively. 

\section{Spectral characteristics of the impedance: Qualitative analysis }

Figure~4(a) shows the functions $G(\pi\,X)$ and $H(\pi\,X$) (inset). As seen, $G(\pi\,X)= G(\omega\,t^i/2)$
is negative  in certain frequency ranges $\Omega_m^0 < \omega < \Omega_m^1$
with  $\Omega_m^0$ and   $\Omega_m^1$ obeying  the equations $G(\Omega_m^0t^i/2) = 0$
and $ G(\Omega_m^1t^i/2)=0$, respectively.
The function $G(\pi\,X)$
exhibits minima at $X = X_m$, where $m = 1, 2,3,..$ and $X_1 \simeq 0.632$,  $X_2 \simeq 1.632$, $X_3 \simeq 2.632$,...
In these minima $G(\pi\,X_m) = G_m < 0$. The minima correspond to the characteristic TT  frequencies $\Omega_m^N/2\pi = X_m/t^i = X_mv_W/2l$, where  $\Omega_m^0 < \Omega_m^N < \Omega_m^1$.
Figure~4(b) shows the frequencies $\Omega_m^N/2\pi$ as functions of the i-region length $2l$. 
For sufficiently short i-regions ($2l \lesssim 1~\mu$m), $\Omega_m^0/2\pi$, $\Omega_m^N/2\pi$, and 
$\Omega_m^1/2\pi$ are in the THz range.

Since in certain  frequency ranges  $G(\pi\,X)= G(\omega\,t^i/2) < 0$, 
  the impedance real part Re~$Z_{\omega}$  given by Eq.~(13)
can be negative provided that the net d- and f-region resistance  $2F/\sigma^d$ is sufficiently small.

The function $H(\pi\,X)$ can change the sign if $d[{\mathcal J}_0(\pi\,X) \sin(\pi\,X)]/dX|_{X=0} > \pi\xi^i$.
This is possible if $\xi^i < 1$. 
If $H(\omega\,t^i/2)$ is close to zero, and  the contribution of d-regions inductance, reflected by the first term on the right-hand side of Eq.~(14),  is small, the impedance imaginary part Im~$Z_{\omega}$ can turn to zero.
This can occur at $\omega = \Omega_m^R > \Omega_m^0$. 
The frequencies $\Omega_m^R$ can be referred to
as the TT resonant frequencies.
Depending on the structure parameters,
$\Omega_m^R$ can be larger than $\Omega_m^0$, but smaller than $\Omega_m^1$. In this case, Im~$Z_{\omega}$ turns to zero in the frequency range in which Re~$Z_{\omega} < 0$. In this case, the self-excitation of the THz oscillations in the GPIN-TTD might be possible~\cite{46}. Thus, the GPIN-TTD impedance frequency dependence is different whether $\Omega_m^R$ falls
into the interval $(\Omega_m^0, \Omega_m^1)$ or not (with  $\Omega_m^0 < \Omega_m^N < \Omega_m^1$).

For example, setting $2l = 0.3 - 0.5~\mu$m, $L = 0.5~\mu$m, $\mu = {\overline \mu}= 1.22T$, and ${\overline \Phi} = 100$~mV, we obtain $\sigma^i = (3.6 - 2.8)$~ps$^{-1}$, min$(2F/\sigma^d) = 0.069$~ps,  $\Lambda = 2.89 -2.36$, and $\omega/2\pi = \Omega^N_1/2\pi \simeq 2.11 - 1.26$ THz with $G_m \simeq - 0.1$ [see Fig.~4(a)].
In this case, the inequality $\xi^i < 1$ is satisfied for $k_{eff} < 1.18 - 2.125$. At $\overline \Phi = 400$~mV,
instead of the latter inequality one obtains $k_{eff} < 2.36 - 4.25$.
For  longer i- and d-regions [$2l = 1.0~\mu$m, $L = 1.0~\mu$m, when $\sigma^i = (2.0 - 4.0)$~ps$^{-1}$, min$(2F/\sigma^d) = 0.0927$~ps, and $\Omega^N_1/2\pi \simeq 0.632$~THz], the condition in question is more liberal: 
$k < 3.09 -6.18$.
However, since the ratio $|G_m|/\Omega^2_m$ steeply drops with increasing index $m$, 
the above condition can be realized for $m= 1$ only.

As follows from Eq.~(14), at $\omega = \Omega_{TT}$, where

 \begin{eqnarray}\label{eq17}
\Omega_{TT} = \sqrt{\frac{\sigma^d\nu}{2c^i}} = \frac{1}{\sqrt{{\mathcal L}^dc^i}},
\end{eqnarray}
the impedance imaginary part turns to zero.

Comparing the characteristic frequencies $\Omega_{PL}$, $\Omega_{TT}$, and $\Omega^N_m$, for their ratio we obtain

 \begin{eqnarray}\label{eq18}
 \frac{\Omega_{PL}}{\Omega_{TT}} = \sqrt{\frac{8\pi^2c^i}{\kappa}} = \sqrt{4\ln \Lambda},  
\end{eqnarray}

\begin{eqnarray}\label{eq19}
\frac{\Omega^N_1}{\Omega_{TT}} =\frac{X_1\hbar\,v_W}{e(2l)}\sqrt{\frac{8\pi^3\,L\kappa\,c^i}{\mu}} =\frac{X_1\hbar\,v_W}{e(2l)}\sqrt{\frac{4\pi\,L\kappa\Lambda}{\mu}}.
\end{eqnarray}
Setting as in the above estimates,   $2l = (0.3 -0.5)~\mu$m, $L = 0.5~\mu$m ($\Lambda = 2.89 - 2.364$), $\kappa =1.5 - 2.5$, and $\mu = {\overline \mu}= 1.22T$,
we obtain $\Omega_{PL}/\Omega_{TT} \simeq 3.51 - 3.03$ and $\Omega^N_1/2\pi = 2.11 - 1.26$~THz with
$\Omega^N_1/\Omega_{TT} \simeq 1.01 - 0.55$ (for $\kappa_{eff} = 1.5$) and
$\Omega^N_1/\Omega_{TT} \simeq 1.3 - 0.71$ (for $\kappa_{eff} = 2.5)$.

The estimates demonstrate that at realistic device parameters, Re~$Z_{\omega}$ can be negative in the THz frequency range. Moreover, in some cases, at the frequencies corresponding to  Re~$Z_{\omega} < 0$, the impedance imaginary part Im~$Z_{\omega}$ can turn to zero. As seen from these estimates, the disregarding of the PL response assumed above is justified up to the frequencies $\omega \sim \Omega^N_1$ (at least for not too short i-region when $\Omega_1 \ll \Omega_{PL}$). The ratio $\Omega_1/\Omega_{PL}$ is small because the ungated d-regions capacitance
is smaller than the i-region capacitance.

\section{Spectral characteristics of the impedance: numerical calculations}

Figures~5 - 7 show the dependences of the GPIN-TTD impedance real and imaginary parts, Re $Z_{\omega}$ and Im $Z_{\omega}$,
on the signal frequency $f =\omega/2\pi$ calculated numerically using Eq.~(12) [or Eqs.~(13) - (16)] for different parameters 
$2l$, $L$ (or $\eta =\nu\,L/v_W$), and $k_{eff}$, and for different values of DC voltage drop, ${\overline \Phi}$, across the i-region. It is assumed that $T = 300$~K and $\nu = 1$~ps. Other parameters are taken from Table I.

As seen from these figures, Re~$Z_{\omega}$ of the GPIN-TTDs with  the chosen parameters is negative in  certain
frequency ranges around the frequency $\Omega_1^N/2\pi$, i.e., the frequencies corresponding to the TT fundamental resonances. At some parameters and elevated voltage ${\overline \Phi}$, Re~$Z_{\omega} < 0$ also near the second resonance with the frequency about $\Omega_2^N$ [see Figs.~5(b) and 6(b)]. The negativity of the impedance real part can be accompanied by the zero imaginary part. As follows from Figs.~5 - 7, this  occurs at the frequency close to the fundamental TT resonance, but not in the case at the second resonance [see Figs.~5(b) and 6(b)]. 
To illustrate this, Figs.~8(a) and 8(b) show the plots taken from Figs.~5(a), 6(a) and 5(b), 6(b) corresponding to $\kappa_{eff} = 2$ and the frequencies
$\Omega_1^N/2\pi = 0.63$~THz and $\Omega_1^N/2\pi = 0.84$~THz, respectively.

\begin{figure*}[t] \centering
\includegraphics[width=13.0cm]{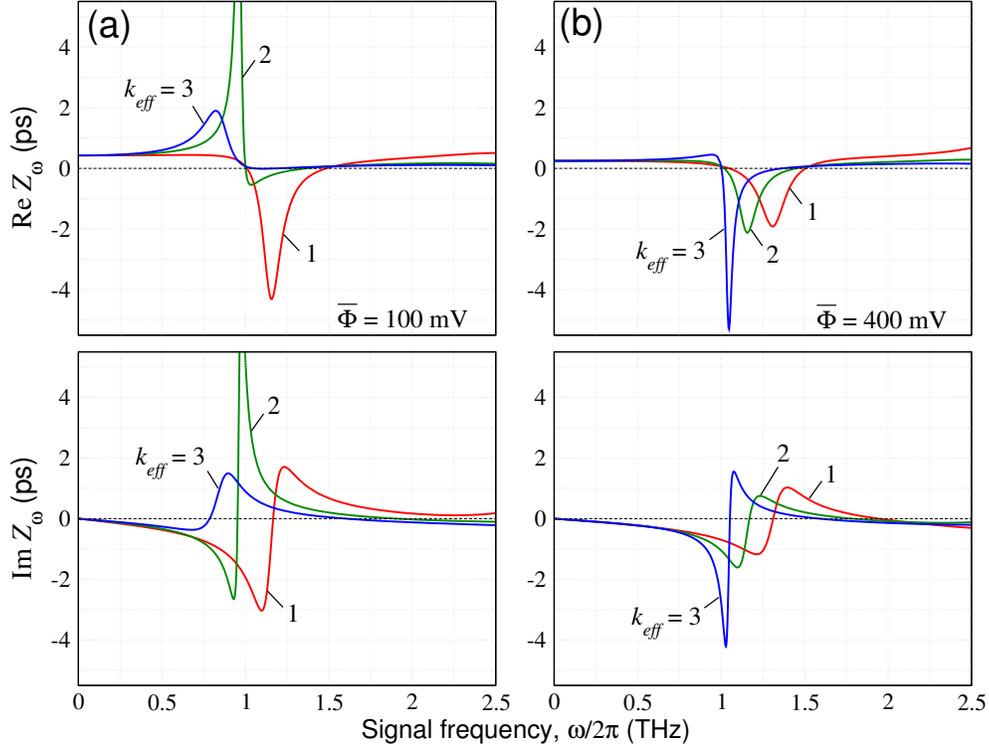}
\caption{
The same as in Figs.~5 and 6, but 
for  
the GPIN-TTD with $2l = 0.5~\mu$m and $L = 0.5~\mu$m
 for different $k_{eff}$ at (a) ${\overline \Phi} = 100$~mV and (b)
  ${\overline \Phi} = 400$~mV ($\Omega_1^N/2\pi  \simeq 1.26$~THz).
} 
\label{F7ab}
\end{figure*}
\begin{figure*}[t] \centering
\includegraphics[width=13.0cm]{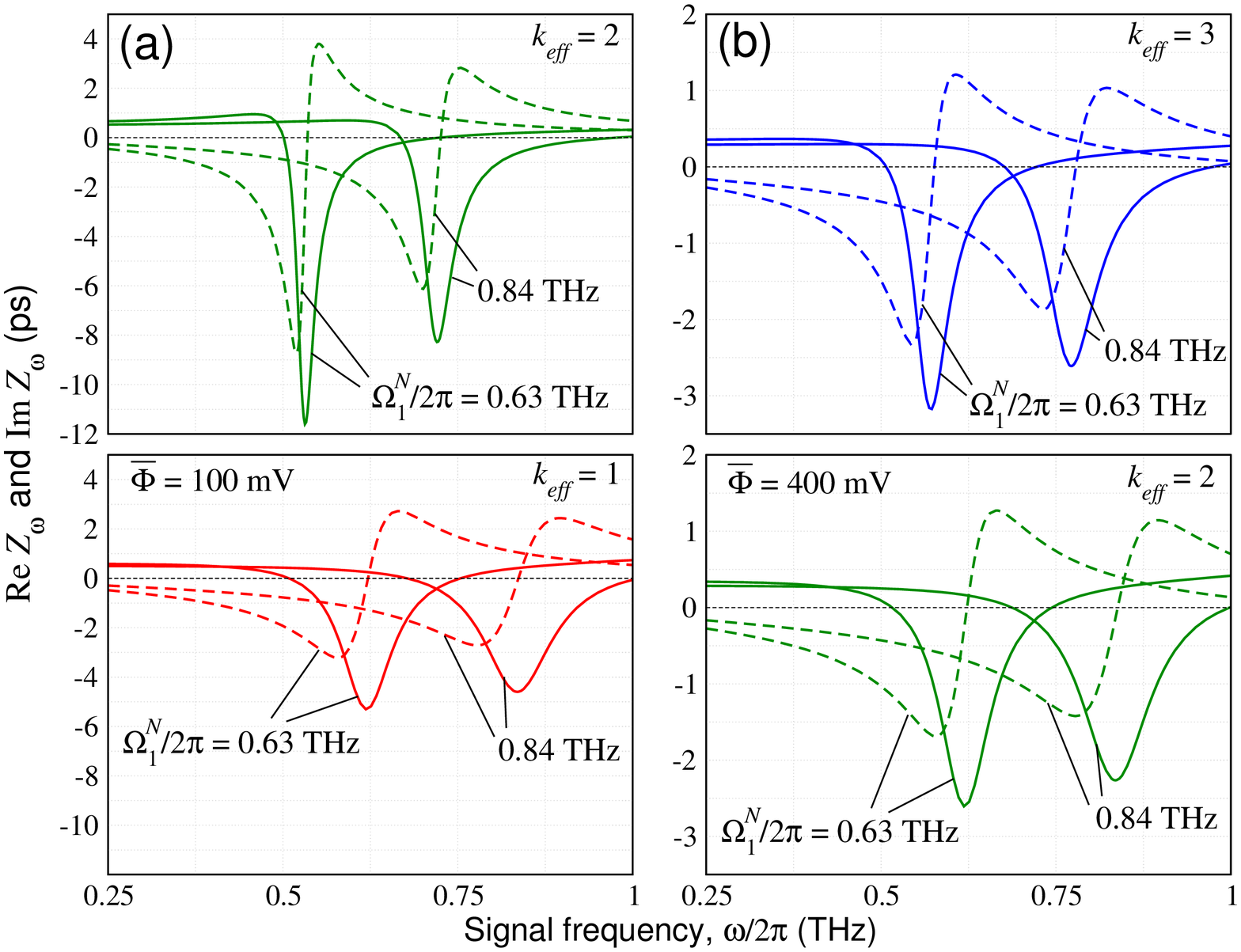}
\caption{The frequency dependences of the GPIN-TTD impedance real (solid lines) and imaginary parts (dashed lines) for $\kappa_{eff} = 1, 2,$ and 3 and other  structural parameters corresponding to
the TT characteristic frequencies $\Omega_1^N/2\pi = 0.63$~THz and $\Omega_1^N/2\pi = 0.84$~THz (shown also in Figs.~5 and 6):
 (a) ${\overline \Phi} = 100$~mV and (b)
  ${\overline \Phi} = 400$~mV.
} 
\label{F8ab}
\end{figure*}

\section{Absorption/amplification and generation of  incident THz radiation}

When the i-region Re~$\sigma^i_{\omega} > 0$, so that Re~$Z_{\omega}$ is positive, the GPIN-TTD absorbs the incident radiation. If  Re~$\sigma^i_{\omega} < 0$ in a certain frequency range, Re~$Z_{\omega}$ can be
negative in this range. In the latter case, the GPIN-TTD can serve as an active region of the devices generating
electromagnetic radiation, particularly, in the THz range.
Thus,
the GPIN-TTD exhibiting Re~$Z_{\omega} < 0$ and 
supplied with an antenna can be the THz radiation source. The pertinent conditions are Im~$Z_{\omega} = 0$ and Re~$Z_{\omega}H + R^A = 0$, where $R^A$ is the antenna radiation resistance (see, for example,~\cite{46}) and $H$ is the GPIN-TTD width. 
Assuming that in line with Figs.~5 - 8, at the frequency corresponding to Im~$Z_{\omega} = 0$, Re~$Z_{\omega} \simeq -2.5$~ps ($\sim 2.25\, \Omega\cdot$cm), $H = 10^3~\mu$m, for the periodic GPIN-TTDs with $N= 10$ sections (periods), one obtains
Re~$NZ_{\omega}/H \simeq  -225~\Omega$. This implies that the impedance  matching requires an antenna with sufficiently
large radiation resistance like those considered in~\cite{47,48}.

Sufficiently long GPIN-TTD without a special antenna (having only the highly conducting side contacts)
with the impedance matched to the free space impedance $Z = 4\pi/c \simeq (4\pi/3)10^{-10}$~s/cm (in the Gaussian units or $Z =120\pi~\Omega \simeq 377~\Omega)$, can effectively emit the THz radiation. For the parameters as in the above estimate, this requires the number of the structure sections  $N \simeq 16 - 17$.

The GPIN-TTD with Re~$Z_{\omega} < 0$ can amplify the incident THz radiation (the absorption coefficient is negative). In this situation, the THz source can consist of such a GPIN-TTD placed into a Fabri-Perot cavity.
The  coefficient, $A_{\omega}$, of the normally incident THz radiation  absorption in the
GPIN-TTD channel 
is expressed via the impedance
${\tilde Z}_{\omega}$ as (see, for example,~\cite{49,50,51})

\begin{eqnarray}\label{eq20}
A_{\omega} = 1 - \biggl|\frac{{\tilde Z}_{\omega}/ Z- 1}
{{\tilde Z}_{\omega}/Z+ 1}\biggr|^2 \nonumber\\
=
\frac{4  {\rm Re}~{\tilde Z}_{\omega}/ Z}
{({\rm Re}~{\tilde Z}_{\omega}/ Z + 1)^2 + ({\rm Im}~{\tilde Z}_{\omega}/ Z)^2},
\end{eqnarray} 
where
${\tilde Z}_{\omega} = Z_{\omega}\sqrt{\kappa_{eff}}/D$ and $D = 4l+ 2L$ is the section length.

First of all, we estimate $A_{\omega}$ for the frequency corresponding to Im~$Z_{\omega} = 0$. Setting as in the above estimate Re~$Z_{\omega} \simeq -(2.5 -5.0)$~ps, $D =  (2.0 - 4.0)~\mu$m, $\kappa_{eff} = 1 - 3$, and accounting that $|{\tilde Z}_{\omega}/Z| \gg 1$, we obtain from Eq.~(20)

\begin{eqnarray}\label{eq21}
A_{\omega} \simeq  \frac{4Z}{{\rm Re}{\tilde Z}_{\omega}} \simeq -(5 -23)~\%.
\end{eqnarray}

Figure~9 shows the absorption coefficient $A_{\omega}$ versus the frequency of the incident THz radiation $\omega/2\pi$
calculated using Eqs.~(12) and (20) for the GPIN-TTD parameters corresponding to the plots in Fig.~8. 
As seen, the absorption coefficient $A_{\omega}$ exhibits a resonant plasmonic peak at the frequencies
close to the TT characteristic frequency $\Omega_1^N$, i.e, to the fundamental TT resonance.
The estimate given by Eq.~(21) is in line with the results of numerical calculations shown in Fig.~9.\\

 \begin{figure*}[t] \centering
\includegraphics[width=13.0cm]{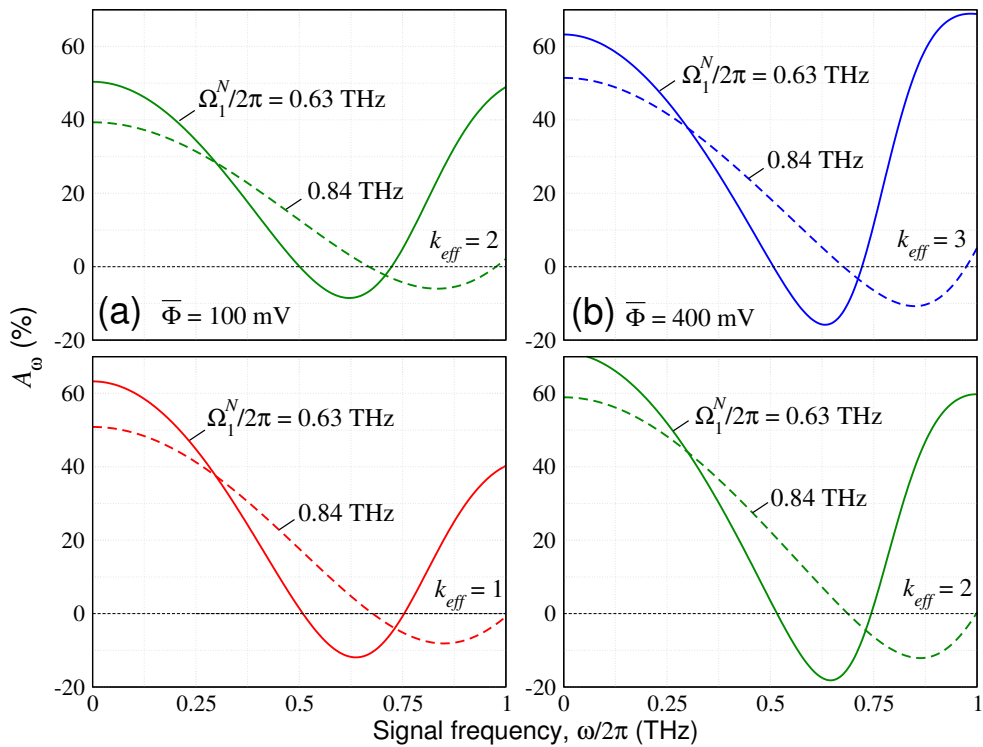}
\caption{The frequency dependences of the absorption coefficient, $A_{\omega}$, of the  GPIN-TTDs
with the parameters corresponding to different values of the $\kappa_{eff}$ and   $\Omega_1^N = 0.63$~THz - solid lines
and  $\Omega_1^N = 0.84$~THz - dashed lines
:
(a) ${\Phi} = 100$~mV and (b)  ${\overline \Phi} = 400$~mV.
} 
\label{F9ab}
\end{figure*}

\section{Comments}
(i) At elevated bias voltages, the potential drop across the i-region can be so large that $e{\Phi} > \hbar\omega_0$,
where $\hbar\omega_0\simeq 200$~meV is the optical phonon energy in GLs.
In this situation, a fraction of the electrons and holes generated in the i-region spontaneously emits
the optical phonons. This leads to a loss of these carriers energy and  momentum. As a result,
the directed velocity of the carrier emitted the optical phonons and the average directed velocity
$<v_x>$ become  smaller than $\pm v_W$. Hence, the optical phonon emission at elevated voltages can
result in an increase of the carrier TT and a decrease in the characteristic TT frequency that leads to
a red shift of the impedance and absorption coefficient peaks, although a rather moderate [about of $(3 - 5)\,\%$]
for the parameters  corresponding to the plots in Figs.~6(b) and 7(b) (see Appendix B).

(ii) The realization of the negative impedance in the GPIN-TTD and the amplification of the THz radiation
passing its plane, associated with the TT resonances, require low or moderate values of the  THz dielectric constants of the substrate and the top/passivating layer.  Among such materials one can mention 
crystalline hBN as a material for the substrate  with a thin passivation layer ($\kappa_{eff} \simeq 3$)~\cite{52}
as well as polyimide, porous BN/polyimide composite, and amorphous HBN ($\kappa \simeq 1.16$)~\cite{53,54}. 
It is interesting that the GPIN-TTDs on some of such substrates can be used for flexible THz sources (see~\cite{55}).

(iii) When the DC  voltage drop across the i-region ${\overline \Phi}$ is relatively large, the space charge  of the tunnel-generated carriers [disregarded above, in particular, in Eq.~(4)] can play some role modifying the DC potential spatial distribution~\cite{28}. This, in turn, leads
to a variation (to some increase as shown by numerical simulations~\cite{28}) of the tunnel current. The space-charge effects in the structures under consideration are characterized by the parameter $\gamma_{SC} = [(e^2/4\pi^2\kappa_{eff}\hbar^{3/2}\,v_W^{3/2})\sqrt{2le{\overline \Phi}}]$~\cite{28}.
The effects in question become crucial when $\gamma_{SC} \gtrsim 1$. 
The parameter $\gamma_{SC}$ reflects the ratio of the electric field formed by the hole and  electron charges in the i-regions and the electric field created by the DC voltage between the p- and n-regions. Setting $2l = 0.5 - 0.75~\mu$m and $\kappa_{eff} = 2 - 3 $
for ${\overline \Phi} = (100 - 400)$~mV, we obtain $\gamma_{SC} \simeq 0.055- 0.101$. This implies that in the GPIN-TTDs at elevated voltages (${\overline \Phi} \gtrsim 400$~mV), the space-charge effects can be pronounced.
 However, such effects  increase  the differential conductance
$\sigma^i$ and, therefore,  increase of the absolute value of the negative dynamic conductance. The latter is useful for reinforcing  the THz radiation
amplification and generation. The GPIN-TTD structures with multiple non-Bernal (twisted) GLs (instead of a single GL in the devices considered above) also can be used as the THz sources. 
In these multiple-GPIN-TTDs the negative conductances of the i-regions of each GL add up, while the net geometrical capacitance is close to that for single-GL structures.
Due the net space-charge in the parallel i-regions, their net conductance  can be rather marked affecting the device performance. 
In general, the  multiple-GPIN-TTDs can surpass the single-GPIN-TTDs, although their analysis
 requires a separate study.

                                                                                                                                                                                    (iv) It is interesting to compare the  amplification coefficient estimate  given by Eq.~(21) with the                                                                                                                                                                                    interband amplification coefficient of the GL with the optical or injection pumping exhibiting the interband population inversion~\cite{56,57,58,59,60}. The amplification coefficient maximum of the latter is equal max~($A_{\omega}^{int}) \simeq \pi\,e^2/\hbar\,c\sqrt{\kappa_{eff}}$.
This yields max~($A_{\omega}^{int}) \simeq (1.3 - 2.3)\,\%$, i.e., markedly smaller than the GPIN-TTD maximum amplification coefficient.
In contrast to  the pumped GLs, which can amplify the THz radiation in a fairly wide frequency range,
the GPIN-TTDs exhibit the amplification at frequencies close to the TT resonances.                                                                                                                                                                                 
The GPIN-TTD structures comprise  the reverse- and forward-biased p-i-n junctions. Similar situation occurs in the dual-gated
structures~\cite{15}.
The p-i-n junctions of the both types can exhibit the
THz radiation amplification but associated with different physical mechanisms. 
The co-existence of  these mechanisms can reinforce the amplification effect.

                                                                                                                                                                                  (v) The  values of the  amplification coefficient, -$A_{\omega}$,  demonstrated above                                                                                                                                                                                     correspond to                                                                                                                                                                           nonoptimal match                                                                                                                                                                                                         of the GPIN-TTD impedance and the free space impedance. 
The device parameters used in the above calculation correspond                                                                                                                                                                                 to large values of  $|{\rm Re} {\tilde Z}_{\omega}/Z| \propto|{\rm Re}~Z_{\omega}|$. Properly choosing the conductance of the d- and f-regions (by varying the doping level and the d-region length), one can, in principle,  achieve the condition ${\rm Re}~{\tilde Z\omega}/Z \sim 1$ (with   Re~$Z_{\omega} < 0$). One can expect that such an optimization can provide a fairly large amplification coefficient and even to achieve $A_{\omega} > 1$.

                                                                                                                                                                           (vi) Above, in particular, in Eq.~(1), we neglected the resistance of the side contacts. This is justified by the possibility to form very low resistance contact to GLs                                                                                                                                                                               ~\cite{61,62,63,64,65}, as low as  $0.0045~\Omega\cdot$cm~\cite{63} and                                                                                                                                                     $0.02~\Omega\cdot$cm~\cite{65} for  Au contacts. Hence, the contact resistance can be much smaller (by two or more orders of magnitude) than the impedance of the forward-biased n-i-p junctions $Z_{\omega}^f \sim (4 - 5)~\Omega\cdot$cm.

(vii) When the PL and TT frequencies are comparable, the interaction of the pertinent resonances can add complexity
to the pattern of the phenomena under consideration. 
In contrast to the situation analyzed above, this can happen primarily in the  GPIN-TTDs  with the {\it gated}
p- and n-regions, which are beyond the scope of this paper.

\section{ Conclusions}
We  proposed GPIN-TTDs based on periodic ungated GL structures with the p-i-n junctions exhibiting the TT effects associated with the holes and electrons generated in the reverse-biased i-regions due to
the Zener-Klein interband tunneling.
The proposed GPIN-TTDs
 can be used for the amplification and generation of the THz radiation.
Their operation is associated with the TT resonances in certain ranges of frequencies resulting in the negative dynamic conductance of the i-regions. 
The TT resonant frequencies are determined by the i-region capacitance and the kinetic inductance of the tunnel-generated carriers in the i-regions as well as  the carriers in the doped p- and n-regions.
Since the GPIN-TTDs can demonstrate rather strong  amplification of the THz signals,
the THz sources based on the GPIN-TTDs coupled to an antenna (or being used as distributed coupling antenna structure), forming the ring circuit configuration, or 
 placed within a Fabri-Perot cavity,  
 can markedly surpass the emitters using the interband negative conductance caused by
optical or injection pumping.

\section*{Acknowledgments}

The Japan Society for Promotion of Science (KAKENHI Grants $\#$ 21H04546 and $\#$ 20K20349), Japan; RIEC Nation-Wide Collaborative Research Project $\#$. R04/A10; the US Office of Scientific, Research Contract
N00001435, (Project Monitor Dr. Ken Goretta).



\section*{Appendix A.} 

\setcounter{equation}{0}
\renewcommand{\theequation} {A\arabic{equation}}

\subsection*{The impedance of forward-biased n-p junctions}

The current via the undoped regions between the p- and n-regions  forward-biased lateral p-i-n junctions in the GPIN-TTDs (we refer to this region as the f-regions)
comprises the tunneling and thermionic components. The former is characterized by the differential conductance similar to that for the reversed-bias n-i-p junction given by Eq.~(4), but with the voltage, ${\overline \Phi}^f$, markedly smaller than ${\overline \Phi}$. Due to this, the  tunneling component is much smaller than the thermionic component.
Considering that  the electrons and holes injected from the n- and p-regions   into the  p- and n-
regions, respectively, have  kinetic energies exceeding $\mu - e{\overline \Phi}^f$ and taking into account the angular
spread of the electron and hole velocities,
the density of the thermionic current via the forward-bias p-i-n junction in the  GPIN-TTDs under consideration
can be presented as

\begin{eqnarray}\label{eqA1}
{\overline J}^{f} \simeq \frac{4eT^2}{\hbar^2v_W}\biggl(\frac{\mu}{T}\ln2 +\frac{\pi^2}{12}\biggr)\exp\biggl(-\frac{\mu}{T}\biggr)\nonumber\\
\times \biggl[\exp\biggl(\frac{e{\overline \Phi}^{f}}{T}\biggr) -1\biggr].
\end{eqnarray}
Considering the smallness of  the voltage drop across the forward-biased n-p junction ${\overline \Phi}^{f}$,
and disregarding the kinetic inductance of this junction, from Eq.~(A1) we obtain the following formula for the junction impedance:

\begin{eqnarray}\label{eqA2}
Z_{\omega}^f \simeq Z_{0}^f =
 \frac{\pi^2\hbar^2v_W\exp(\mu/T)}{4e^2(\mu\ln2 + \pi^2T/12)}.
\end{eqnarray}

In particular, accounting for Eq.~(A2), we obtain

%

\begin{eqnarray}\label{eqA3}
{\rm Re}(2 Z_{\omega}^d + Z_{\omega}^f)
\simeq \frac{2}{\sigma^d}\biggl[1 + \biggl(\frac{\pi\,v_W}{8L\nu}\biggr)\frac{\mu\exp(\mu/T)}{(\mu\ln2 + \pi^2T/12)}\biggr].\qquad
\end{eqnarray} 

\section*{Appendix B. 
}
\setcounter{equation}{0}
\renewcommand{\theequation} {B\arabic{equation}}

\subsection*{Role of the optical phonons emission}

Assuming for simplicity that the optical phonon emission makes the carrier
distribution semi-isotropic, the average directed carrier velocity after the emission can be set as  $ v_x \simeq \pm v_W/2$.
Hence, for the carriers in the i-regions where their energy ${\mathcal E} < \hbar\omega_0$ the average velocity is equal to
$\pm v_W$  and for the carriers in the region where ${\mathcal E} > \hbar\omega_0$ this velocity is equal to $\pm v_W/2$.
Considering that the characteristic time, $\tau_0$, of the spontaneous optical phonon emission is finite,
a portion of the carriers having the energy ${\mathcal E} > \hbar\omega_0$ can continue to propagate with the velocity
$\pm v_W$. Summarizing all this, we arrive at the following rough estimate for $<v_x>$:

\begin{eqnarray}\label{eqB1}
\frac{<v_x>}{v_W} \simeq \frac{1}{2}\biggl(1 + \frac{l_0}{l}\biggr)
 + 
\frac{v_W\tau_0}{4l} [1 - e^{-2(l-l_0)/v_W\tau_0}],
 \end{eqnarray} 
where
\begin{eqnarray}\label{eqB2}
\frac{l_0}{l} = \frac{e{\overline \Phi}-\hbar\omega_0}{e{\overline \Phi}} \cdot \Theta\biggl(\frac{e{\overline \Phi}-\hbar\omega_0}{e{\overline \Phi}}\biggr).
\end{eqnarray} 
When $v_W\tau_0 > 2l$, at $e{\overline \Phi} > \hbar\omega_0$ from Eqs.~(B1) and (B2) for the average velocity $<v_x>/v_W$ and, consequently,  for  the average transit time 
$t^i_0 = 2l/<v_x> = t^iv_W/<v_x>$ 
one obtains

\begin{eqnarray}\label{eqB3}
\frac{<v_x>}{v_W} \simeq 1  - \frac{(l-l_0)^2}{2lv_W\tau_0}, \, t^i_0 \simeq t^i\biggl[1 + \frac{(l-l_0)^2}{2lv_W\tau_0}\biggr].
\end{eqnarray} 

Assuming that $\tau_0 \simeq 1$~ps (see, for example,~\cite{66,67}), for $2l = (0.5 - 0.75)~\mu$m and ${\overline \Phi} = 400$~mV [this corresponds to the plots in Figs.~6(b) and 7(b)] and using Eqs.~(B2) and (B3), we find $(t^i_0-t^i)/t^i \simeq (3.1 - 4.7)\,\%$.   

A low density of states near the Dirac point leads to a decrease of the optical emission probability in comparison with $\tau_0^{-1}$.  Anisotropy of the carrier scattering on the optical phonons results in somewhat larger values of the directed carrier velocity compared to $v_W/2$. This implies that the above estimates provide a conservative value of the average velocity and, hence, an overrated  TT value.

\end{document}